
%
%
\input harvmac
%
%
%
%
%
%
%
%
%
\newif\ifdraft

\noblackbox
\catcode`\@=11
\newif\iffrontpage
%
\ifx\answ\bigans
\def\titleft{\titsm}
\magnification=1200\baselineskip=15pt plus 2pt minus 1pt
%
\advance\hoffset by-0.075truein
\hsize=6.15truein\vsize=600.truept\hsbody=\hsize\hstitle=\hsize
\else\let\lr=L
\def\titleft{\titla}
\magnification=1000\baselineskip=14pt plus 2pt minus 1pt
%
\vsize=6.5truein
\hstitle=8truein\hsbody=4.75truein
\fullhsize=10truein\hsize=\hsbody
\fi
\parskip=4pt plus 10pt minus 4pt

\font\titla=cmr10 scaled\magstep3
\font\tenmss=cmss10
\font\absmss=cmss10 scaled\magstep1
\newfam\mssfam
\font\footrm=cmr8  \font\footrms=cmr5
\font\footrmss=cmr5   \font\footi=cmmi8
\font\footis=cmmi5   \font\footiss=cmmi5
\font\footsy=cmsy8   \font\footsys=cmsy5
\font\footsyss=cmsy5   \font\footbf=cmbx8
\font\footmss=cmss8
\def\footfont{\def\rm{\fam0\footrm}
\textfont0=\footrm \scriptfont0=\footrms
\scriptscriptfont0=\footrmss
\textfont1=\footi \scriptfont1=\footis
\scriptscriptfont1=\footiss
\textfont2=\footsy \scriptfont2=\footsys
\scriptscriptfont2=\footsyss
\textfont\itfam=\footi \def\it{\fam\itfam\footi}
\textfont\mssfam=\footmss \def\mss{\fam\mssfam\footmss}
\textfont\bffam=\footbf \def\bf{\fam\bffam\footbf} \rm}
\def\tenpoint{\def\rm{\fam0\tenrm}
\textfont0=\tenrm \scriptfont0=\sevenrm
\scriptscriptfont0=\fiverm
\textfont1=\teni  \scriptfont1=\seveni
\scriptscriptfont1=\fivei
\textfont2=\tensy \scriptfont2=\sevensy
\scriptscriptfont2=\fivesy
\textfont\itfam=\tenit \def\it{\fam\itfam\tenit}
\textfont\mssfam=\tenmss \def\mss{\fam\mssfam\tenmss}
\textfont\bffam=\tenbf \def\bf{\fam\bffam\tenbf} \rm}
\ifx\answ\bigans\def\abstractfont{\tenpoint}\else
\def\abstractfont{\def\rm{\fam0\absrm}
\textfont0=\absrm \scriptfont0=\absrms
\scriptscriptfont0=\absrmss
\textfont1=\absi \scriptfont1=\absis
\scriptscriptfont1=\absiss
\textfont2=\abssy \scriptfont2=\abssys
\scriptscriptfont2=\abssyss
\textfont\itfam=\bigit \def\it{\fam\itfam\bigit}
\textfont\mssfam=\absmss \def\mss{\fam\mssfam\absmss}
\textfont\bffam=\absbf \def\bf{\fam\bffam\absbf}\rm}\fi
%
\def\f@@t{\baselineskip10pt\lineskip0pt\lineskiplimit0pt
\bgroup\aftergroup\@foot\let\next}
\setbox\strutbox=\hbox{\vrule height 8.pt depth 3.5pt width\z@}
\def\vfootnote#1{\insert\footins\bgroup
\baselineskip10pt\footfont
\interlinepenalty=\interfootnotelinepenalty
\floatingpenalty=20000
\splittopskip=\ht\strutbox \boxmaxdepth=\dp\strutbox
\leftskip=24pt \rightskip=\z@skip
\parindent=12pt \parfillskip=0pt plus 1fil
\spaceskip=\z@skip \xspaceskip=\z@skip
\Textindent{$#1$}\footstrut\futurelet\next\fo@t}
\def\Textindent#1{\noindent\llap{#1\enspace}\ignorespaces}
\def\footnote#1{\attach{#1}\vfootnote{#1}}%

\def\foot{\attach\footsymbolgen\vfootnote{\footsymbol}}
\let\footsymbol=\star
\newcount\lastf@@t           \lastf@@t=-1
\newcount\footsymbolcount    \footsymbolcount=0
\def\footsymbolgen{\relax\footsym
\global\lastf@@t=\pageno\footsymbol}
\def\footsym{\ifnum\footsymbolcount<0
\global\footsymbolcount=0\fi
{\iffrontpage \else \advance\lastf@@t by 1 \fi
\ifnum\lastf@@t<\pageno \global\footsymbolcount=0
\else \global\advance\footsymbolcount by 1 \fi }
\ifcase\footsymbolcount \fd@f\star\or
\fd@f\dagger\or \fd@f\ast\or
\fd@f\ddagger\or \fd@f\natural\or
\fd@f\diamond\or \fd@f\bullet\or
\fd@f\nabla\else \fd@f\dagger
\global\footsymbolcount=0 \fi }
\def\fd@f#1{\xdef\footsymbol{#1}}
\def\space@ver#1{\let\@sf=\empty \ifmmode #1\else \ifhmode
\edef\@sf{\spacefactor=\the\spacefactor}
\unskip${}#1$\relax\fi\fi}
\def\attach#1{\space@ver{\strut^{\mkern 2mu #1} }\@sf\ }
%
\newif\ifnref
\def\rrr#1#2{\relax\ifnref\nref#1{#2}\else\ref#1{#2}\fi}

\nreffalse
\def\refout{\listrefs}
%
\def\eqn#1{\xdef #1{(\secsym\the\meqno)}
\writedef{#1\leftbracket#1}%
\global\advance\meqno by1\eqno#1\eqlabeL#1}
\def\eqnalign#1{\xdef #1{(\secsym\the\meqno)}
\writedef{#1\leftbracket#1}%
\global\advance\meqno by1#1\eqlabeL{#1}}
%
\def\chap#1{\newsec{#1}}
\def\chapter#1{\chap{#1}}
\def\sect#1{\subsec{#1}}
\def\section#1{\sect{#1}}
\def\\{\ifnum\lastpenalty=-10000\relax
\else\hfil\penalty-10000\fi\ignorespaces}
\def\note#1{\leavevmode%
\edef\@@marginsf{\spacefactor=\the\spacefactor\relax}%
\ifdraft\strut\vadjust{%
\hbox to0pt{\hskip\hsize\hskip.05in%
\vbox to0pt{\vskip-\dp\strutbox%
\sevenrm\baselineskip=10pt plus 1pt minus 1pt%
\ifx\answ\bigans\hsize=.9in\else\hsize=.4in\fi%
\tolerance=5000 \hbadness=5000%
\leftskip=0pt \rightskip=0pt \everypar={}%
\raggedright\parskip=0pt \parindent=0pt%
\vskip-\ht\strutbox\noindent\strut#1\par%
\vss}\hss}}\fi\@@marginsf\kern-.01cm}
\def\titlepage{%
\frontpagetrue\nopagenumbers\abstractfont%
\hsize=\hstitle\rightline{\vbox{\baselineskip=10pt%
{\abstractfont\pubnum}}}\pageno=0}
\frontpagefalse
\def\pubnum{}
\def\pdate{\number\month/\number\yearltd}
\def\makefootline{\iffrontpage\vskip .27truein
\line{\the\footline}
\vskip -.1truein\line{\pdate\hfil}
\else\vskip.5cm\line{\hss \tenrm $-$ \folio\ $-$ \hss}\fi}
\def\title#1{\vskip .7truecm\titlestyle{\titleft #1}}
\def\titlestyle#1{\par\begingroup \interlinepenalty=9999
\leftskip=0.02\hsize plus 0.23\hsize minus 0.02\hsize
\rightskip=\leftskip \parfillskip=0pt
\hyphenpenalty=9000 \exhyphenpenalty=9000
\tolerance=9999 \pretolerance=9000
\spaceskip=0.333em \xspaceskip=0.5em
\noindent #1\par\endgroup }
\def\autskip{\ifx\answ\bigans\vskip.5truecm\else\vskip.1cm\fi}
\def\author#1{\vskip .7in \centerline{#1}}
\def\andauthor#1{\autskip
\centerline{\it and} \autskip\centerline{#1}}
\def\address#1{\ifx\answ\bigans\vskip.2truecm
\else\vskip.1cm\fi{\it \centerline{#1}}}
\def\abstract#1{\vskip .5in\vfil\centerline
{\bf Abstract}\penalty1000
{{\smallskip\ifx\answ\bigans\leftskip 2pc \rightskip 2pc
\else\leftskip 5pc \rightskip 5pc\fi
\noindent\abstractfont \baselineskip=12pt
{#1} \smallskip}}
\penalty-1000}
\def\endpage{\tenpoint\supereject\global\hsize=\hsbody%
\frontpagefalse\footline={\hss\tenrm\folio\hss}}
%
\def\CERN{\address{CERN, Geneva, Switzerland}}
\def\inbar{\vrule height1.5ex width.4pt depth0pt}
\def\IC{\relax\,\hbox{$\inbar\kern-.3em{\mss C}$}}
\def\IF{\relax{\rm I\kern-.18em F}}
\def\IH{\relax{\rm I\kern-.18em H}}
\def\II{\relax{\rm I\kern-.17em I}}
\def\IN{\relax{\rm I\kern-.18em N}}
\def\IP{\relax{\rm I\kern-.18em P}}
\def\IQ{\relax\,\hbox{$\inbar\kern-.3em{\rm Q}$}}
\def\IR{\relax{\rm I\kern-.18em R}}
\def\ZZ{\relax{\hbox{\mss Z\kern-.42em Z}}}
\def\nup#1({Nucl.\ Phys.\ $\us {B#1}$\ (}
\def\plt#1({Phys.\ Lett.\ $\us  {B#1}$\ (}
\def\plb#1({Phys.\ Lett.\ $\us  {#1B}$\ (}
\def\cmp#1({Comm.\ Math.\ Phys.\ $\us  {#1}$\ (}
\def\prp#1({Phys.\ Rep.\ $\us  {#1}$\ (}
\def\prl#1({Phys.\ Rev.\ Lett.\ $\us  {#1}$\ (}
\def\prv#1({Phys.\ Rev. $\us  {#1}$\ (}
\def\und#1({            $\us  {#1}$\ (}
\def\tit#1,{{\it #1},\ }
%

\def\tilde{\widetilde}
\def\bar{\overline}
\def\us#1{\bf{#1}}

\def\Coe#1.#2.{{#1\over #2}}

\def\coe#1.#2.{\relax{\textstyle {#1 \over #2}}\displaystyle}

\def\notin{\hbox{{$\in$}\kern-.51em\hbox{/}}}

\catcode`\@=12


\def\LOUIS{\rrr\LOUIS{J. Louis, {\it
         ``Non-harmonic gauge coupling constants in supersymmetry
         and superstring theory'',} preprint SLAC-PUB-5527 (1991);
         V. Kaplunovsky and J. Louis, as quoted in J. Louis,
         SLAC-PUB-5527 (1991).}}

\def\LEH{\rrr\LEH{J. Lehner, Discontinuous groups and automorphic
functions, ed. The American Mathematical Society (1964).}}

\def\DHVW{\rrr\DHVW{L. Dixon, J. Harvey, C.~Vafa and E.~Witten,
         \nup261 (1985) 651;
        \nup274 (1986) 285.}}

\def\DKLA{\rrr\DKLA{L. Dixon, V. Kaplunovsky and J. Louis, \nup329 (1990)
            27;
            V. Kaplunovsky and J. Louis, paper to appear?}}

\def\IBLU{\rrr\IBLU{L. Ib\'a\~nez and D. L\"ust, \nup382 (1992) 305.}}

\def\BALO{\rrr\BALO{D. Bailin, S. Gandhi and A. Love, \plt269 (1991)
293, \plt275 (1992) 55; D. Bailin and A. Love, \plt288 (1992) 263.}}

\def\GEP{\rrr\GEP{D. Gepner, \nup296 (1998) 757.}}

\def\ILLT{\rrr\ILLT{ L.E. Ib\'a\~nez, W. Lerche, D. L\"ust and
S. Theisen, \nup352 (1991) 435.}}

\def\CREMMER{\rrr\CREMMER{E. Cremmer, S. Ferrara, L. Girardello and
          A. Van Proeyen, \nup212 (1983) 413.}}

\def\DFKZ{\rrr\DFKZ{
J.P. Derendinger, S. Ferrara, C. Kounnas and F. Zwirner,
         \nup372 (1992) 145,
          \plt271 (1991) 307.}}

\def\LAUERA{\rrr\LAUERA{
J. Lauer, J. Mas and H.P. Nilles, \nup351 (1991) 353.}}

\def\LAUER{\rrr\LAUER{
J. Lauer, J. Mas and H.P. Nilles, \plt226 (1989) 251;
            W. Lerche, D. L\"ust and N.P. Warner, \plt231 (1989) 417;
     E.J. Chun, J. Mas, J. Lauer and H.-P. Nilles, \plt233 (1989) 141.}}

\def\MODULAR{\rrr\MODULAR
        {R.~Dijkgraaf, E.~Verlinde and H.~Verlinde, \cmp115 (1988) 649;
        {\it ``On moduli spaces of conformal field theories with $c\geq
        1$'',} preprint THU-87/30;
           A. Shapere and F. Wilczek, \nup320 (1989) 669;
                 S. Ferrara,
         D. L\"ust, A. Shapere and S. Theisen, \plt225 (1989) 363;
 S. Ferrara, D. L\"ust and S. Theisen, \plt233 (1989) 147;
                      M. Cvetic, A. Font, L.E.
           Ib\'a\~nez, D. L\"ust and F. Quevedo, \nup361 (1991) 194.}}

\def\DHS{\rrr\DHS{M. Dine, P. Huet and N. Seiberg, \nup322 (1989) 301.}}

\def\DUAL{\rrr\DUAL{K. Kikkawa and M. Yamasaki, \plb149 (1984) 357;
      N. Sakai and I. Senda, Progr. Theor. Phys. {\bf 75} (1986) 692.}}

\def\CAOV{\rrr\CAOV{G. Lopes Cardoso and B. Ovrut,
   \nup369 (1992) 351.}}

\def\LT{\rrr\LT{D. L\"ust and S. Theisen, \nup302 (1988) 499.}}

\def\SCHEWA{\rrr\SCHEWA{
A.N. Schellekens and N.P. Warner, \nup308
(1988) 397, \nup 313 (1989) 41.}}

\def\KW{\rrr\KW{
L. Krauss and F. Wilczek, Phys.Rev.Lett. {\bf 62} (1989) 1221.}}

\def\FIQ{\rrr\FIQ{
A. Font, L.E. Ib\'a\~nez and F. Quevedo, \plt224 (1989) 79;
A. Font, L.E. Ib\'a\~nez, M. Mondragon, F. Quevedo and G.G. Ross,
\plt227 (1989) 34.}}

\def\IR{\rrr\IR{L.E. Ib\'a\~nez and G.G. Ross, \plt260 (1991) 291;
\nup368 (1992) 3; L.E. Ib\'a\~nez, CERN-TH.6662 (1992).}}

\def\OFFSET{\hoffset=6.pt\voffset=40.pt}

\OFFSET

%


\def\pubnum{
\hbox{CERN-TH.6737/92}\hbox{FTUAM/92/40}}

\def\pdate{December 1992}

\titlepage
\title
{A Comment on Duality Transformations and (Discrete) Gauge Symmetries
in Four-Dimensional Strings}
\vskip-.8cm
\author{{\bf Luis Ib\'a\~nez}}
\centerline{{\it Departamento de F\'isica Teorica}}
\centerline{{\it Universidad Aut\'onoma de Madrid, 28049 Madrid, Spain}}
\vskip.2cm
\andauthor{{\bf Dieter L\"ust}                        }
\CERN
\vskip-2.8 cm
\abstract{ We discuss the relationship between target space modular
invariance and discrete gauge symmetries in four-dimensional
orbifold-like strings. First we derive the modular transformation
properties of various string vertex operators of the massless
string fields.
Then we find that for supersymmetric compactifications
the action of the duality elements, leaving invariant   the
multicritical points, corresponds to a combination of finite K\"ahler
and gauge transformations. However, those finite gauge transformations
are not elements of a remnant discrete gauge symmetry. We suggest
that, at least
in the case of Gepner models corresponding to tensor products
of identical minimal models, the duality element leaving
invariant the multicritical point corresponds to the ${\bf Z}_{k+2}$
symmetry of any of the minimal $N=2$ models appearing in the
tensor product.
}

\vskip2.5cm
\noindent CERN-TH.6737\endpage

In four-dimensional strings, moduli fields in general have
non-vanishing charges under local gauge symmetries. This implies
that parts of the local gauge symmetries are spontaneously broken
at generic points in moduli space. However usually, a {\it discrete}
gauge symmetry \KW\
               survives this spontaneous symmetry breakdown.
On the other hand,
target space duality transformations \DUAL, in
particular target space modular transformations \MODULAR,
act non-trivially
on many light four-dimensional string fields. Moreover, particular
elements of the duality group act  with simple, constant phases
on the fields. Therefore duality symmetries act similarly to
discrete symmetries. (We will show that they act like $R$-symmetries.)
The similarities between discrete gauge symmetries and
{\it some} duality symmetries naively suggest considering these
duality symmetries as just another example of discrete gauge symmetries.
However, this naive suggestion has to be qualified. For example,
it is well known that the {\it massless} spectrum of orbifold models
typically has duality anomalies \DFKZ ,\LOUIS ,\CAOV ,\IBLU\
whereas the enhanced gauge
(discrete or continuous) symmetries are anomaly-free (at least for
(2,2) models).
Thus things
are not so simple and, although indeed there is a connection between
duality and discrete gauge transformations, the identification is not
straightforward.

The intention of this letter is to clarify
the relation between target space modular transformations,
broken gauge symmetries and discrete gauge groups in four-dimensional
string models. As a specific example we consider the ${\bf Z}_3$
orbifold \DHVW.\foot{For previous discussions on the relation
between duality symmetries and broken gauge symmetries
in the ${\bf Z}_3$ orbifold see \DHS,\ILLT.}
However the discussion can be easily extended
to other models.

Let us determine the transformation properties of the
vertex operators of various fields under target space modular
transformations $T\rightarrow{aT-ib\over icT+d}$ for the ${\bf Z}_3$
orbifold. As explained in \ILLT, these transformation rules
can be derived from the action of the modular transformations on the
momentum and winding numbers and from
the subsequent action on the Narain
lattice vectors. Specifically, consider the left (right) moving
complex coordinate $X^{+i}_{L(R)}$ ($i=1,2,3$) which is associated with
the $i^{th}$ 2-dimensional torus of the ${\bf Z}_3$ orbifold.
For a general $PSL(2,{\bf Z})$ transformation one finds
$$X^{+i}_L(\bar z)\rightarrow\lambda_i
\Biggl\lbrack{-ic_i\bar T_i+d_i\over ic_iT_i+d_i}\Biggr\rbrack
^{-1/2}X^{+i}_L(\bar z),\qquad
X^{+i}_R(z)\rightarrow\lambda_i
\Biggl\lbrack{-ic_i\bar T_i+d_i\over ic_iT_i+d_i}\Biggr\rbrack
^{1/2}X^{+i}_R(z),\eqn\xlr
$$
where $\lambda_i$ is a $T_i$-independent phase that depends on the
parameters $c_i,d_i$: e.g. $\lambda=\rho=e^{2\pi i/3}$ for $c=d=1$
($ST$ transformation).

Next consider the corresponding right-moving world sheet fermions
$\psi^{+i}_R$ with conformal dimension $h_R=1/2$.
Their transformation behaviour can be deduced from
the requirement \DHS\ that the right-moving world sheet supersymmetry
commutes with the target space modular transformations.
This requirement follows from the fact that the action of the
right-moving supercurrent connects equivalent (picture-changed)
physical string states.
The right-moving
world sheet supercurrent has the form
$$S_R(z)=\sum_{i=1}^3(\psi^{+i}_R\partial X^{-i}_R+\psi^{-i}_R\partial
X^{+i}_R)(z).\eqn\supcur
$$
Demanding $S_R$ to be invariant
under modular transformations, one obtains
$$
\psi^{+i}_R(z)\rightarrow\lambda_i
\Biggl\lbrack{-ic_i\bar T_i+d_i\over ic_iT_i+d_i}\Biggr\rbrack
^{1/2}\psi^{+i}_R(z).\eqn\psir
$$
For many purposes it is very convenient to bosonize the fermions
$\psi^{+i}_R$: $\psi^{+i}_R(z)=e^{iH^i_R(z)}$.
Then modular transformations
act on the two-dimensional bosons $H_i$ as
$$
H^i_R(z)\rightarrow H^i_R(z)-i\log\lambda_i
\Biggl\lbrack{-ic_i\bar T_i+d_i\over ic_iT_i+d_i}\Biggr\rbrack
^{1/2}.\eqn\hr
$$

Next consider the bosonic twist field vertex operators
$\sigma^i_{\alpha_i}(\bar z,z)$
of conformal dimension $h_L=h_R=1/9$.
Each twist field vertex is associated with one of the three fixed points
$\alpha_i$ ($\alpha_i=1,\dots ,3$)
of each complex $i$.
The
twists fields $\sigma_{\alpha_i}$
of different fixed points transform into
linear combinations under target space modular transformations \LAUER.
In addition there is a common field-dependent phase factor, which
was determined in ref.\LAUERA.
In total
one gets
$$
\sigma^i_{\alpha_i}(\bar z,z)\rightarrow
\Biggl\lbrack{-ic_i\bar T_i+d_i\over ic_iT_i+d_i}\Biggr\rbrack
^{1/6}A_{\alpha_i\beta_i}\sigma^i_{\beta_i}(\bar z,z),\eqn\sig
$$
Specifically for the $ST$ transformation the matrix $A$
has the form
$$A={i\over\sqrt 3}
\pmatrix{\rho&\bar\rho
&\bar\rho\cr \rho&\rho&1\cr\rho&1&\rho}
{}.
\eqn\matrixst
$$

Now we are ready to examine the modular transformation properties
of some specific massless string states. First the vertex
operator associated with the modulus $T_i$ has the form (we will only
show the internal parts of the vertex operators)
$$\phi_T^i(\bar z,z)
\sim \bar\partial X^{-i}_L(\bar z)\partial X^{+i}_R(z).\eqn\tfield
$$
Thus $\phi_T^i$ transforms under $PSL(2,{\bf Z})$ as
$$\phi_T^i(\bar z,z)\rightarrow
\Biggl\lbrack{-ic_i\bar T_i+d_i\over ic_iT_i+d_i}\Biggr\rbrack
\phi_T^i(\bar z,z).\eqn\ttrans
$$

Next consider the scalar fields $\phi^i_{1,2}$ whose mass depends
on the background parameters $T_i$. They become massless
at the critical point $T_c=-ie^{2\pi i/3}$. The corresponding
vertex
operators look like
$$\phi^i_{1,2}(\bar z,z)\sim V_{1,2}(\bar z,z,T_i)
\partial X^{+i}_R(z);\eqn\phiv
$$
$V_{1,2}(\bar z,z,T_i)$ is a conformal field that depends on the
background $T_i$ (see \DHS,\ILLT\ for details).
Thus they transform as
$$\phi^i_{1,2}(\bar z,z)\rightarrow\lambda_i
\Biggl\lbrack{-ic_i\bar T_i+d_i\over ic_iT_i+d_i}\Biggr\rbrack
^{1/2}\phi^i_{1,2}(\bar z,z).\eqn\phitr
$$

In the untwisted sector of the ${\bf Z}_3$ orbifold, each complex
plane is associated with three fields, which transform
under the ${\underline{27}}$
representation of $E_6$. Their vertex operators
are obtained by action with the left-moving supercurrent on
the vertex operators of the moduli $T_i$:

$$\phi^{ij}_{27U}(\bar z,z)\sim\psi^{+j}_L(\bar z)\partial
X^{+i}_R(z).\eqn\untw
$$
(We left out some part of the vertex operator associated with the gauge
group
$E_6$.) Thus we obtain that the fields
$\phi^{ij}_{27U}$ transform in the same way as the fields
$\phi^i_{1,2}$
(see eq.\phitr).

Next let us consider the fields in the twisted sector of the ${\bf
Z}_3$
orbifold. First we have 27 fields, associated with the 27 fixed points
of the ${\bf Z}_3$ orbifold, which transform like the
                ${\underline{27}}$ representation
of $E_6$. Their vertex operators are built by the product of the
three
bosonic twist fields $\sigma^i_{\alpha_i}$.
In addition the twist also acts on
the
left and right moving world sheet fermions. In total one obtains
$$\phi_{27T}^{\alpha_1\alpha_2\alpha_3}(\bar z,z)
\sim\prod_{i=1}^3e^{iH^i_L(\bar z)
/3}\sigma^i_{\alpha_i}(\bar z,z)e^{iH^i_R(z)/3}.\eqn\twisma
$$
Thus these fields transform as
$$
\phi_{27T}^{\alpha_1\alpha_2\alpha_3}(\bar z,z)\rightarrow\prod_{i=1}^3
\bar\lambda_i^{2/3}
\Biggl\lbrack{-ic_i\bar T_i+d_i\over ic_iT_i+d_i}\Biggr\rbrack
^{1/3}A_{\alpha_1\beta_1}A_{\alpha_2\beta_2}A_{\alpha_3\beta_3}
\phi_{27T}^{\beta_1\beta_2\beta_3}(\bar z,z).\eqn\twistr
$$

Finally there are 81 $E_6$ singlets in the twisted sector. They contain
left moving twisted oscillators. (Some linear combinations
of them, the twisted moduli,
are obtained by acting with $S_L$ on $\phi_{27T}$.) Their vertex
operators
look like
$$\phi^{i\alpha_1\alpha_2\alpha_3}_{1T}(\bar z,z)\sim
e^{-i2H^i_L(\bar z)/3}e^{iH^j_L(\bar z)/3}e^{iH^k_L(\bar z)
/3}\tau^i_{\alpha_i}(\bar z,z)
\sigma^j_{\alpha_j}(\bar z,z)
\sigma^k_{\alpha_k}(\bar z,z)\prod_{l=1}^3e^{iH^l_R(z)/3}.\eqn\twissin$$
Here $i\neq j\neq k$ and $\tau^i$ is an excited twist field obtained by
the
operator product of $\partial X^{-i}_L$ and $\sigma^i$.

Then $\phi^i_{1T}$ transforms as
$$\eqalign{
\phi^{i\alpha_1\alpha_2\alpha_3}(\bar z,z)
_{1T}&\rightarrow
\bar\lambda_i^{5/3}
\Biggl\lbrack{-ic_i\bar T_i+d_i\over ic_iT_i+d_i}\Biggr\rbrack^{5/6}
\bar\lambda_j^{2/3}
\Biggl\lbrack{-ic_j\bar T_j+d_j\over ic_jT_j+d_j}\Biggr\rbrack^{1/3}
\cr & \bar\lambda_k^{2/3}
\Biggl\lbrack{-ic_k\bar T_k+d_k\over ic_kT_k+d_k}\Biggr\rbrack^{1/3}
A_{\alpha_1\beta_1}A_{\alpha_2\beta_2}A_{\alpha_3\beta_3}
\phi^{i\beta_1\beta_2\beta_3}_{1T}(\bar z,z),\cr}\eqn\twistrs
$$

So far we considered the transformation rules of the scalar
components
of the chiral superfields. To obtain the action of the
modular group on the corresponding space-time fermions, one has to
examine the vertex operator of the space-time supercharge. Its
internal
part has the form
$$Q(z)\sim \prod_{i=1}^3e^{-iH^i_R(z)/2}.\eqn\superc
$$
Therefore $Q$ transforms as
$$Q(z)\rightarrow\prod_{i=1}^3\bar\lambda_i^{1/2}
\Biggl\lbrack{-ic_i\bar T_i+d_i\over ic_iT_i+d_i}\Biggr\rbrack
^{-1/4}Q(z).\eqn\qtrs$$
Thus we recognize that the fermions get an additional phase under
modular transformations. Therefore target space modular
transformations
act like an $R$ symmetry. From the field theory point of view, we will
identify this additional
phase as a K\"ahler transformation.

To summarize this analysis, the modular transformation behaviour of
the
massless string fields has the following form (up to the
field-independent phase):
$$\phi_s
\rightarrow\prod_{i=1}^3
\Biggl\lbrack{-ic_i\bar T_i+d_i\over ic_iT_i+d_i}\Biggr\rbrack
^{-n^i_s/2}\phi_s.\eqn\qtrsgen
$$
$n^i_s$ is called
the modular weight vector of $\phi_s$. Specifically, comparing with our
previous results, we have
$$\eqalign{\phi_T^i:&\qquad \vec n=-2\vec e_i\cr
\phi^i_{1,2},\phi^{ij}_{27U}:&\qquad \vec n=-\vec e_i\cr
\phi_{27T}:&\qquad \vec n=(-2/3,-2/3,-2/3)\cr
\phi^i_{1T}:&\qquad \vec n=(-2/3,-2/3,-2/3)-\vec e_i\cr
Q:&\qquad \vec n=(1/2,1/2,1/2).\cr}\eqn\summ
$$
(The $\vec e_i$ are the 3-dimensional unit-vectors.)
Notice that the above numbers for $\vec n$ correspond to
the modular weights of massless states discussed in ref. \IBLU .

Now let us consider the particular element, $\gamma=ST$, of the target
space
modular group acting as $T_i\rightarrow{1\over T_i-i}$, i.e. $a_i=0$,
$b_i=-1$,
$c_i=d_i=1$. This transformation leaves the critical point
$T_c=-ie^{2\pi i/3}$ invariant.
Moreover, one can find a basis in which
the $ST$ transformation acts diagonally on
particular linear combinations of the twist fields of the form
$\sum_\alpha c_\alpha \sigma^i_\alpha$.
Specifically, the $ST$ charges of these particular linear combinations
of twist fields
are given by the eigenvalues of the matrix in eq.\matrixst:
$A'={\rm diag}(\bar\rho,\bar\rho,\rho)$.
Thus, at the critical point, all fields
transform under the $ST$ modular transformation as
$$\phi_s\rightarrow e^{2\pi
iQ^i_{ST}}\phi_s.\eqn\dualch
$$
We call $Q^i_{ST}$ the {\it duality charge} of each field.
With ${-i\bar T_c+1\over iT_c+1}=e^{4\pi i/3}$ one obtains the
following duality charges:
$$\eqalign{\phi_T^i:&\qquad Q_{ST}^i=2/3\cr
\phi^i_{1,2},\phi^{ij}_{27U}:&\qquad Q_{ST}^i=2/3\cr
\phi_{27T}^{1,2}:&\qquad Q_{ST}^i=0\cr
\phi_{27T}^3:&\qquad Q_{ST}^i=2/3\cr
\phi^{i1,2}_{1T}:&\qquad Q_{ST}^i=0\cr
\phi^{i3}_{1T}:&\qquad Q_{ST}^i=2/3\cr
}\eqn\dualchf
$$

On the other hand, all fields are characterized by certain $U(1)$
gauge charges. First each complex plane is associated with an
enhanced
$U(1)_1^i\times U(1)_2^i$ gauge group which is left unbroken at the
critical
point $T_i=T_c$.
As discussed in refs.\LT,\SCHEWA, at the critical point
it is possible to `rebosonize' those parts of the
vertex operators that involve the torus coordinates $X^{\pm i}_{L,R}$.
In this new, so-called covariant lattice
basis the left-moving (and also the right-moving) part
of the vertex operators of
all fields we have considered so far can be written as
$$V_L(\bar z)
\sim\prod_{i=1}^3\exp\biggl(i(Q_1^iY_1^i(\bar z)+Q_2^iY_2^i(\bar z)
)\biggr),\eqn
\rebos
$$
where $Y_{1,2}^i$ are the covariant lattice coordinates. The conformal
dimensions are just given by $h_L={1\over 2}\sum_{i=1}^3\biggl(
(Q_1^i)^2+(Q_2^i)^2\biggr)$.
For example, the left-moving torus coordinates can be written as
$i\bar\partial X^{\pm}_L(\bar z)={1/\sqrt 3}\sum_{\vec\alpha}
\exp(\pm i\vec\alpha\cdot \vec Y(\bar z))$, where $\pm\vec\alpha$ are
the six root vectors of $SU(3)$ with $\vec\alpha^2=2$.
The gauge
bosons of the enhanced $U(1)^6$ gauge group, which are massless for
$T=T_c$,
correspond just to
$\bar\partial Y_{1,2}^i(\bar z)$.
In the twisted sectors, the fields with definite $U(1)$ charges
just correspond to those linear combinations of twist fields,
on which $ST$ acts in a diagonal way.

Let us consider the particular $U(1)^i$ subgroup of $U(1)_1^i\times
U(1)_2^i$,
with
charge $Q_{U(1)}^i$, defined by the following linear combination
$$Q_{U(1)}^i={\sqrt 2\over 3}Q_1^i+{2\over\sqrt 6}Q_2^i.\eqn\lincom
$$
In fact, these charges $Q_1^i$ and $Q_2^i$ are just given
by the charges that appear in the vertex operators eq.\rebos\
in the covariant lattice basis.
The various fields have the following $U(1)$ gauge charges \LT:

$$\eqalign{\phi_T^i,\phi^i_{1,2}
:&\qquad Q_1^i=\sqrt2,\qquad Q_2^i=0,\qquad Q_{U(1)}^i=2/3\cr
&\qquad Q_1^i=-{1\over\sqrt2},\qquad Q_2^i=\pm{3\over\sqrt6}
,\qquad Q_{U(1)}^i=2/3\cr
\phi^{ij}_{27U}:&\qquad Q_1^i=0,\qquad Q_2^i=0,\qquad Q_{U(1)}^i=0\cr
\phi_{27T}^1:&\qquad Q_1^i={1\over
3\sqrt2},\qquad Q_2^i=-{1\over\sqrt6},\qquad Q_{U(1)}^i=-2/9\cr
\phi_{27T}^2:&\qquad Q_1^i=-{\sqrt2\over
3},\qquad Q_2^i=0,\qquad Q_{U(1)}^i=-2/9\cr
\phi_{27T}^3:&\qquad Q_1^i={1\over
3\sqrt2},\qquad Q_2^i={1\over\sqrt6},\qquad Q_{U(1)}^i=4/9\cr
\phi^1_{1T}:&\qquad Q_1^i=-{2\over 3\sqrt2},\qquad Q_2^i={2\over\sqrt6}
,\qquad Q_{U(1)}^i=4/9\cr
\phi^2_{1T}:&\qquad Q_1^i={2\sqrt2\over 3},\qquad Q_2^i=0
,\qquad Q_{U(1)}^i=4/9\cr
\phi^3_{1T}:&\qquad Q_1^i=-{2\over 3\sqrt2},\qquad Q_2^i=-{2\over\sqrt6}
, \qquad Q_{U(1)}^i=1/9\cr
}\eqn\gaugech
$$
Second each complex plane is associated with a $U(1)$ holonomy charge
$Q_{Hol}$. This charge  is a linear combination
of the superconformal $U(1)$ inside $E_6$ and of the two Cartan
subalgebra
$U(1)'s$ of $SU(3)$: $Q^{superconf}=\sum_{i=1}^3Q^i_{Hol}$,
$Q^{SU(3)_1}={1\over\sqrt2}(Q^1_{Hol}-Q^2_{Hol})$,
$Q^{SU(3)_2}={1\over\sqrt6}(Q^1_{Hol}+Q^2_{Hol}-2Q^3_{Hol})$;
$Q^i_{Hol}$ is determined by the left-moving
world sheet fermions $\psi^i_L$. Specifically, the left-moving
fermionic part of the vertex operators has the form
$e^{i{3\over2}Q^i_{Hol}H^i_L}$. Thus we obtain:

$$\eqalign{T_i:&\qquad Q_{Hol}^i=0\cr
\phi^i_{1,2}:&\qquad Q_{Hol}^i=0\cr
\phi^{ij}_{27U}:&\qquad Q_{Hol}^i=2/3\cr
\phi_{27T}:&\qquad Q_{Hol}^i=2/9\cr
\phi^i_{1T}:&\qquad Q_{Hol}^i=-4/9.\cr}
\eqn\holcharge
$$

Now comparing eqs.\dualchf,\gaugech\ and\holcharge, we recognized
that
the charges obey  the following relation
$$Q_{ST}^i=Q^i_{U(1)}+Q^i_{Hol}.\eqn\relat
$$
For the fermions, an additional phase
is involved.
Thus an $ST$ duality transformation  acts like a linear combination
of an enhanced $U(1)$ gauge transformation and a $U(1)$ gauge
holonomy
transformation. Moreover, the $ST$ duality transformation  acts
like an $R$-symmetry.
Equation \relat\ becomes clear when the
left-moving supercurrent $S_L$ is investigated. It
has charges $Q_{U(1)}=2/3$, $Q_{Hol}=-2/3$. Specifically,
the vertex operator
of $S_L$ in the covariant lattice basis has the form $S_L\sim\sum_{i=1}^3
\lbrack \exp\biggl(i\sqrt2 Y_1^i\biggr)
+\exp\biggl(i(-{1\over\sqrt2}Y_1^i+{3\over\sqrt6}Y_2^i)\biggr)
+\exp\biggl(i(-{1\over\sqrt2}Y_1^i-{3\over\sqrt6}Y_2^i)\biggr)\rbrack
\exp(-iH_L^i)$. Then, the requirement that $Q_{ST}=0$ for $S_L$
implies eq.\relat.

It is also interesting to consider the overall target space
modular transformations, i.e. simultaneous transformations
on all $T_i$ with $a_1=a_2=a_3$ etc. With eq.\relat\ it is easy
to see that an overall $ST$ transformation
acts exactly like a linear combination of all $U(1)^9$
gauge transformations.

                       Note that the ${\bf Z}_3$ orbifold can be
equivalently constructed by tensoring together nine
$c=1$, $N=2$ superconformal models. Using this method \GEP,
one always constructs the theory at the multicritical point
$T_i=T_c$ with enhanced $U(1)^9$ gauge symmetry.
In this case, the massless bosonic states of the model may be labelled
by giving the nine $q$-charges of the nine chiral states with
$(l,q,s)=(1,1,0)$ appearing in the tensor product. For example
(see e.g. ref.\FIQ ),
            the untwisted $\underline {27}$'s correspond to
charges $(q_1,q_2,.....,q_9)=$  $(1,1,1;0,0,0;0,0,0)$,
$(0,0,0;1,1,1;0,0,0)$, $(0,0,0;0,0,0;1,1,1)$,.... Twisted
$\underline {27}$'s correspond to states labelled
$({\underline {1,0,0}};{\underline {1,0,0}};{\underline {1,0,0}})$,
where the underlining indicates permutations. We have grouped
the charges in sets of three to explicitly show the correspondence
of each set of three factors with one complex dimension.
The symmetry assignments of each massless state with respect to the
${\bf Z}_{k+2}$
symmetries of the model are obtained by computing the scalar
product of vectors of the form
$\gamma =(\gamma _1,.. ..,\gamma _9)$ ($\gamma _i=$integers)
with the $(q_1,...,q_9)$ vectors of each state using a
diagonal metric with $g_{ij}=-\delta _{ij}/(k+2)$, $i,j=1,...,9$.
In particular, we find that the action in the first complex plane of
the ST generator on the massless bosonic fields of the ${\bf Z}_3$
orbifold (eq.\dualchf) is identical to the symmetry generated by
$\gamma =(1,0,0;0,0,0;0,0,0)$, as the reader may easily check.

The above fact suggests that, at least for Gepner models \GEP\
of the type
$(k=n)^m$ ($m$ identical copies of a $k=n$ minimal model), the
duality generator that leaves invariant the multicritical point
corresponds to the ${\bf Z}_{k+2}$ symmetry generated by
symmetries of the type $\gamma =(1,0,..,0)$.

Now let us discuss the target space modular transformations
and their relation to the $U(1)$ gauge transformations
within the 4-dimensional effective field theory of the orbifold
compactified heterotic string.
The kinetic energies of the moduli $T_i$ and the
chiral ``matter'' fields $A_s$ are determined by
the K\"ahler potential of the following (tree level)
form:
$$
K=-\sum_{i=1}^3\log(T_i+\bar T_i)+\prod_{i=1}^3
(T_i+\bar T_i)^{n^i_s}|A_s|^2.\eqn\kp
$$
Invariance of the matter kinetic energies
under target  space modular transformations requires that
                              the chiral superfields as well as their
bosonic components (we denote them by the same symbol $A_s$) transform
like (up to constant matrices and phases):
$$A_s\rightarrow\prod_{i=1}^3(ic_iT_i+d_i)^{n^i_s}A_s.\eqn\sugratrans
$$
Therefore we call the numbers $n^i_s$ the modular weights of the
matter fields. The modular weights $n^i_s$, i.e. the kinetic
energies of the matter fields, were previously
computed
\DKLA,\IBLU,\BALO\
by comparing string
calculations with the effective Lagrangian,
and for the matter fields of the ${\bf Z}_3$ orbifold
the result is given in eq.\summ.

The transformation behaviour of the fermionic
components of the chiral fields in the supergravity Lagrangian
follows from the action of the target space modular group on the
K\"ahler potential.
Specifically $K$ transforms with a K\"ahler transformation
like
$$K\rightarrow K+\Lambda+\bar\Lambda,\qquad\Lambda=
\sum_{i=1}^3(ic_iT_i+d_i).\eqn\ktrnans
$$
Then it follows \CREMMER\ that the fermions $\psi_{A_s}$ transform with
an additional K\"ahler phase as
$$\psi_{A_s}\rightarrow
e^{{1\over4}(\Lambda-\bar\Lambda)}(ic_iT_i+d_i)^{n^i_s}\psi_{A_s}=
\Biggl\lbrack{-ic_i\bar T_i+d_i\over ic_iT_i+d_i}\Biggr\rbrack
^{-1/4}(ic_iT_i+d_i)^{n^i_s}\psi_{A_s}.\eqn\fermtrans
$$
Likewise, the gauginos $\psi_\lambda$ transform as
$$\psi_\lambda\rightarrow
e^{-{1\over4}(\Lambda-\bar\Lambda)}\psi_\lambda=
\Biggl\lbrack{-ic_i\bar T_i+d_i\over ic_iT_i+d_i}\Biggr\rbrack
^{1/4}\psi_\lambda.\eqn\gaugtrans
$$

It is now easy to show that the transformation rules of the fields $A_s$
in the
supergravity Lagrangian agree with the rules we have
obtained in the string theory by examining the corresponding
vertex operators.
In the string basis, the fields $\phi_s$
are represented by vertex operators,
which create normalized states with canonical kinetic energies.
Therefore to relate the string fields with the supergravity fields,
one has to perform the following non-holomorphic field redefinition
(except for the moduli fields $T_i$, see \ILLT):
$$\phi_s=\prod_{i=1}^3(T_i+\bar T_i)^{n^i_s/2}A_s.\eqn\nonhol
$$
Then, using eq.\sugratrans, we immediately obtain the correct
field-dependent phase eq.\qtrsgen\ of the string vertex operators.
Moreover, the string theory also provides the information
about the field-independent phases and matrices, which cannot
be obtained by considering the effective supergravity Lagrangian.
Analogously, the K\"ahler phase
$e^{{1\over 4}(\Lambda-\bar\Lambda)}$ just corresponds to the non-trivial
modular transformation behaviour, eq.\qtrs, of the space-time
supercharge in the string basis.

Let us now show that also within the effective supergravity
Lagrangian, the $ST$-modular transformations act like
a linear combination of field-independent $U(1)$ gauge transformations
plus, for the fermions, a constant K\"ahler phase on the
chiral fields. To achieve this, one has to perform a
field redefinition to a new supergravity field basis, which
allows to couple the charged chiral fields to the $U(1)$ vector
gauge fields. Specifically, for the moduli fields $T_i$ one has
to perform the following {\it holomorphic} field redefinition \ILLT:
$$\tilde T_i={T_c-T\over\bar T_c+T},\qquad T_c=-ie^{2\pi i/3}.\eqn\uniform
$$
Then $\tilde T_i$ transforms under the $ST$ transformation $T_i
\rightarrow{1\over T_i-i}$ as
$$\tilde T_i\rightarrow e^{4\pi i/3}\tilde T_i.\eqn\tildetrns
$$
Thus $\tilde T_i$ transforms under $ST$ exactly like the vertex operator
$\phi_{T_i}$ at the critical point $T_c$, i.e. $ST$ acts on $\tilde T_i$
like a $U(1)$ gauge transformation.
In fact in the literature about modular functions (see e.g. \LEH)
the variable $\tilde T_i^3$ is nothing else than the {\it uniformizing}
variable around the elliptic fixed point $T_c=-ie^{2i\pi/3}$
of the element $\gamma=ST$. In general, the uniformizing
variables are conveniently used to expand meromorphic
functions $F$ around the fixed points of modular transformations.
A function $F$ is single valued if it can be expressed in terms of
integer powers of $\tilde T_i^3$. This is the so-called
{\it uniformization}.

Similarly, the matter fields have to be redefined as follows
$$\tilde A_s=\biggl\lbrack{\sqrt{T_c+\bar T_c}\over\bar T_c+T}
\biggr\rbrack^{-n^i_s}A_s.\eqn\holredmatt
$$
Then the matter fields transform under $ST$ with a constant phase
like displayed in eqs.\dualch\ and \dualchf.
For the fermions similar field redefinitions can be performed and
one obtains that the fermions transform with an additional
constant phase, which shows that $ST$ acts like an $R$-symmetry
in the supergravity Lagrangian.

The K\"ahler potential in the new field basis has the following
form:
$$\tilde K=-\sum_{i=1}^3\log(1-|\tilde T_i|^2)+\prod_{i=1}^3
(1-|\tilde T_i|^2)^{n^i_s}|\tilde A_s|^2.\eqn\newkp
$$
Note that in this field basis the K\"ahler gauge function is a purely
imaginary number, $\tilde\Lambda=-4i\pi/3$. Therefore $ST$ does
not act on $\tilde K$.

It is obvious that the Lagrangian can now be made locally gauge invariant
under
the $U(1)^9$ gauge symmetry by the gauge covariant replacement
$\bar{\tilde A_s}\rightarrow\bar{\tilde A_s}\exp(\sum_{a=1}^9
Q^a_sV_a)$ (here $\tilde A_s$ also includes $\tilde T_i$), where
$V_a$ are the $U(1)^9$ vector fields and the $Q^a_s$ are the
corresponding charges.
Since the moduli fields $\tilde T_i$ are charged under
$U(1)_1^i\times U(1)_2^i$ (see eq.\gaugech),
non-vanishing vacuum expectation values of these fields spontaneously
break these $U(1)$ symmetries, and the corresponding gauge bosons
become massive. (See refs.\DHS,\ILLT\ for details). (Non-vanishing vacuum
expectation values of
twisted moduli
also break the two further $U(1)$'s, which are linear combinations
of $U(1)^i_{Hol}$. Moreover, in general there are also moduli
that are charged under $E_6$.)
However an inspection of the $U(1)$ charges of the various fields shows
that $U(1)_1^i\times U(1)_2^i$ is not broken completely, but
the Lagrangian is still invariant under a {\it discrete} gauge
symmetry ${\bf Z}_3\times {\bf Z}_3$.
Consider for example the group $U(1)^i$ defined by eq.\lincom, with
charges as displayed in eq.\gaugech. Since all untwisted fields,
including the symmetry-breaking field $\tilde T_i$, have charge 2/3,
whereas all twisted fields have charges of units 1/9, a discrete
${\bf Z}_3$ symmetry remains unbroken. Untwisted fields are neutral
under this discrete gauge symmetry, whereas twisted fields have
${\bf Z}_3$ charges $1/3$, $2/3$. This result has to be compared
with the discrete ${\bf Z}_3$ group generated by the modular
element $ST$. Looking at the $ST$ charges of all fields,
eq.\dualchf,
we see that {\it the $ST$ discrete group cannot be identified
with the discrete gauge group discussed above}, since the
symmetry-breaking field $\tilde T_i$ is not inert under $ST$.

Finally let us mention the interesting possibility  \IR\
that the discrete gauge groups are anomalous.
In this case the corresponding anomaly of the
underlying $U(1)$ must be cancelled by the Green-Schwarz
mechanism, i.e. by a non-trivial gauge transformation
of the universal axion field. Similarly target space duality
transformations, which involve an additional K\"ahler phase
for the fermions, may be anomalous  \DFKZ,\LOUIS,\CAOV,\IBLU .
Then the axion transforms non-trivially under modular transformations.
However note that there is no direct relationship between
anomalous discrete gauge symmetries \IR\ and anomalous target
space modular transformations \DFKZ ,\LOUIS ,\CAOV ,\IBLU .
In fact, for the ${\bf Z}_3$
orbifold, looking at the massless spectrum
target space modular transformations are anomalous,
whereas the enhanced $U(1)$ symmetries and thus the discrete
gauge symmetries are anomaly-free.

\vskip2cm

We thank S. Ferrara  and J. Louis for useful discussions.
The work of D.L. was performed as a Heisenberg fellow.

\refout
\vfill\eject\bye